# Flavour violating bosonic squark decays at LHC

**Keisho Hidaka** [1]

*Department of Physics, Tokyo Gakugei University*
*Koganei, Tokyo 184-8501, Japan*
*E-mail:* `hidaka@u-gakugei.ac.jp`

**Alfred Bartl, Elena Ginina**

*University of Vienna, Faculty of Physics*
*Boltzmanngasse 5, A-1090 Vienna, Austria*

**Helmut Eberl, Walter Majerotto**

*Institut für Hochenergiephysik der ÖAW*
*A-1050 Vienna, Austria*

**Björn Herrmann**

*LAPTh, Université de Savoie, CNRS*
*9 Chemin de Bellevue, F-74941 Annecy-le-Vieux, France*

**Werner Porod**

*Institut für Theoretische Physik und Astrophysik, Universität Würzburg*
*D-97074 Würzburg, Germany*

Quark flavour conserving (QFC) fermionic squark decays, such as $\tilde{t}_{1,2} \to t\, \tilde{\chi}_i^0$, are usually assumed in squark search analyses. Here we study quark flavour violating (QFV) bosonic squark decays, such as $\tilde{u}_2 \to \tilde{u}_1\, h^0/Z^0$, where the mass eigenstates $\tilde{u}_{1,2}$ are mixtures of scharm and stop quarks. We show that the branching ratios of such QFV decays can be very large due to sizable $\tilde{c}_R$-$\tilde{t}_{R/L}$ and $\tilde{t}_R$-$\tilde{t}_L$ mixing effects despite the very strong constraints on the QFV parameters from B meson data. This can result in remarkable QFV signatures with significant rates at LHC (14 TeV), such as $pp \to \tilde{g}\tilde{g}X \to t\,c\,\bar{c}\,\bar{c}\, h^0/Z^0\, E_T^{mis}\, X$ and $pp \to \tilde{g}\tilde{g}X \to t\,t\,\bar{c}\,\bar{c}\, h^0/Z^0\, E_T^{mis}\, X$. The QFV bosonic squark decays can play an important role in the squark and gluino searches at LHC (14 TeV).

---

[1] Presenter

## 1. Introduction

Quark flavour conserving (QFC) fermionic squark decays, such as $\tilde{t}_{1,2} \to t \tilde{\chi}_i^0$ are usually assumed in squark searches. However, squark generation mixing in non-minimal flavour violation (NMFV) can induce quark-flavour violating (QFV) fermionic/bosonic squark decays. In this article based on [1,2] we study the QFV effect on bosonic squark decays at LHC (14 TeV) in the general Minimal Supersymmetric Standard Model (MSSM) with focus on the mixing between scharm ($\tilde{c}$) and stop ($\tilde{t}$) quarks.

## 2. QFV scenarios

We take the scenarios A and C in [1] as our QFV reference scenarios here. Input basic MSSM parameters and output observables (such as sparticle masses and decay branching ratios) in these scenarios are shown in [1]. The main features of these scenarios are (i) large QFC/QFV trilinear couplings of $\tilde{t}_R$-$\tilde{t}_L$-$H_2^0$/$\tilde{c}_R$-$\tilde{t}_L$-$H_2^0$, (ii) sizable scharm-stop mixing, (iii) large mass of either $\tilde{t}_L$ or $\tilde{t}_R$, (iv) large mass of the CP-odd neutral Higgs boson $m_{A^0}$ (= 1500 GeV) and large $\tan\beta$ (= 20). Both scenarios are decoupling-Higgs scenarios with $m_{A^0} >> m_{h^0}$ and hence the lightest Higgs boson $h^0$ is SM-like. These scenarios satisfy all the relevant experimental and theoretical constraints such as (i) those from B meson data, (ii) sparticle mass limits from LHC, (iii) observed Higgs boson mass $m_{h^0} \approx 126$ GeV and (iv) those on the trilinear couplings from vacuum stability conditions. The large top trilinear coupling of $\tilde{t}_R$-$\tilde{t}_L$-$H_2^0$ and the large $m_{\tilde{t}_L}$ or $m_{\tilde{t}_R}$ are required to satisfy $m_{h^0} \approx 126$ GeV. Respecting the LHC limits on squark and gluino masses, we assume a gluino ($\tilde{g}$) mass of about 1 TeV in our scenarios.

## 3. QFV bosonic decays of up-type squarks

In scenario A the lightest up-type squark mass eigenstates $\tilde{u}_{1,2}$ are mainly mixtures of $\tilde{c}_R$ and $\tilde{t}_R$ (and $\tilde{t}_L$). They are lighter than the gluino ($m_{\tilde{g}} = 1141$ GeV). All other squarks are much heavier than the gluino and hence it turns out that the strongly interacting sparticles produced at LHC (14 TeV) are practically only $\tilde{u}_{1,2}$ and $\tilde{g}$ in this scenario. In Table 1 we show the decay branching ratios of $\tilde{u}_{1,2}$ and $\tilde{g}$ in this scenario. We see (i) the QFV bosonic squark decay branching ratio $B(\tilde{u}_2 \to \tilde{u}_1 h^0)$ is very large (~50%), (ii) both $B(\tilde{u}_1 \to c \tilde{\chi}_1^0)$ and $B(\tilde{u}_1 \to t \tilde{\chi}_1^0)$ are very large, which can lead to large QFV effects, and (iii) $B(\tilde{g} \to \tilde{u}_2 c/t) = B(\tilde{g} \to \tilde{u}_2 \bar{c}/\bar{t}) + B(\tilde{g} \to \overline{\tilde{u}}_2 c/t)$ is large (~25%). The very large $B(\tilde{u}_2 \to \tilde{u}_1 h^0)$ is due to the following reason: In scenario A the large $\tilde{c}_R$-$\tilde{t}_R$ mixing induces a large mass splitting between $\tilde{c}_R$ and $\tilde{t}_R$ resulting in two mass eigenstates $\tilde{u}_1$ and $\tilde{u}_2$ with a large mass difference. Here note that $m_{\tilde{t}_R} \approx 750$ GeV and $m_{\tilde{c}_R} \approx 780$ GeV and that $m_{\tilde{u}_1} \approx 605$ GeV and $m_{\tilde{u}_2} \approx 861$ GeV. Moreover, in this scenario with large $\tilde{t}_R$-$\tilde{t}_L$ and $\tilde{c}_R$-$\tilde{t}_R$ mixings and sizable $\tilde{c}_R$-$\tilde{t}_L$ mixing, we have $\tilde{u}_{1,2} \sim \tilde{c}_R + \tilde{t}_R(+\tilde{t}_L)$ and $h^0 \sim \mathrm{Re}(H_2^0)$. Hence the $\tilde{u}_1$-$\tilde{u}_2$-$h^0$ coupling is large due to the large $\tilde{t}_R$-$\tilde{t}_L$-$H_2^0$ coupling $T_{U33}$ (= -2160 GeV). This leads to the very large $B(\tilde{u}_2 \to \tilde{u}_1 h^0)$. The small $B(\tilde{u}_2 \to \tilde{u}_1 Z^0)$ is due to the small $\tilde{t}_L$ components in $\tilde{u}_{1,2}$.

Table 1: Decay branching ratios of $\tilde{u}_{1,2}$ and $\tilde{g}$ in scenario A. The charge conjugated processes have the same branching ratios and are not shown explicitly.

| | | | |
|---|---|---|---|
| $B(\tilde{u}_2 \to \tilde{u}_1 h^0) = 0.47$ | $B(\tilde{u}_2 \to \tilde{u}_1 Z^0) = 0.01$ | $B(\tilde{u}_2 \to c\, \tilde{\chi}_1^0) = 0.43$ | $B(\tilde{u}_2 \to t\, \tilde{\chi}_1^0) = 0.09$ |
| $B(\tilde{u}_1 \to c\, \tilde{\chi}_1^0) = 0.36$ | $B(\tilde{u}_1 \to t\, \tilde{\chi}_1^0) = 0.64$ | | |
| $B(\tilde{g} \to \tilde{u}_2 \bar{c}) = 0.12$ | $B(\tilde{g} \to \tilde{u}_2 \bar{t}) = 0.01$ | $B(\tilde{g} \to \tilde{u}_1 \bar{c}) = 0.09$ | $B(\tilde{g} \to \tilde{u}_1 \bar{t}) = 0.27$ |

In scenario C the lightest up-type squark mass eigenstates $\tilde{u}_{1,2}$ are mainly mixtures of $\tilde{c}_R$ and $\tilde{t}_L$ (and $\tilde{t}_R$). They are lighter than the gluino ($m_{\tilde{g}}$ = 1134 GeV). All other squarks are much heavier than the gluino, except the lightest down-type squark $\tilde{d}_1$, and hence it turns out that the strongly interacting sparticles produced at LHC (14 TeV) are practically only $\tilde{u}_{1,2}$, $\tilde{d}_1$ and $\tilde{g}$ in this scenario. In Table 2 we show decay branching ratios of $\tilde{u}_{1,2}$ and $\tilde{g}$ in this scenario. We see that (i) the QFV bosonic squark decay branching ratios $B(\tilde{u}_2 \to \tilde{u}_1 h^0/Z^0)$ are very large (= 43% / 34%) and (ii) $B(\tilde{g} \to \tilde{u}_2 c/t)$ is large (~ 24%).

Table 2: Decay branching ratios of $\tilde{u}_{1,2}$ and $\tilde{g}$ in scenario C. The charge conjugated processes have the same branching ratios and are not shown explicitly.

| | | | |
|---|---|---|---|
| $B(\tilde{u}_2 \to \tilde{u}_1 h^0) = 0.43$ | $B(\tilde{u}_2 \to \tilde{u}_1 Z^0) = 0.34$ | $B(\tilde{u}_2 \to c\, \tilde{\chi}_1^0) = 0.17$ | $B(\tilde{u}_2 \to t\, \tilde{\chi}_1^0) = 0.06$ |
| $B(\tilde{u}_1 \to c\, \tilde{\chi}_1^0) = 0.96$ | $B(\tilde{u}_1 \to t\, \tilde{\chi}_1^0) = 0.04$ | | |
| $B(\tilde{g} \to \tilde{u}_2 \bar{c}) = 0.04$ | $B(\tilde{g} \to \tilde{u}_2 \bar{t}) = 0.08$ | $B(\tilde{g} \to \tilde{u}_1 \bar{c}) = 0.19$ | $B(\tilde{g} \to \tilde{u}_1 \bar{t}) = 0.05$ |

The very large $B(\tilde{u}_2 \to \tilde{u}_1 h^0)$ is due to the reason similar to that in scenario A: In scenario C we have (i) $m_{\tilde{c}_R} \approx 650$ GeV and $m_{\tilde{t}_L} \approx 780$ GeV and (ii) $m_{\tilde{u}_1} \approx 651$ GeV and $m_{\tilde{u}_2} \approx 800$ GeV. In this scenario with sizable $\tilde{c}_R$-$\tilde{t}_L$ mixing, we have $\tilde{u}_{1,2} \sim \tilde{c}_R + \tilde{t}_L$ and $h^0 \sim \mathrm{Re}(H_2^0)$ and hence the $\tilde{u}_1$-$\tilde{u}_2$-$h^0$ coupling is large due to the large $\tilde{c}_R$-$\tilde{t}_L$-$H_2^0$ coupling $T_{U32}$ (= -500

GeV). This leads to the very large $B(\tilde{u}_2 \to \tilde{u}_1 h^0)$. The very large $B(\tilde{u}_2 \to \tilde{u}_1 Z^0)$ is due to the sizable $\tilde{t}_L$ components in $\tilde{u}_{1,2}$.

## 4. Signatures of QFV bosonic squark decays

In our scenarios, $\tilde{g}$ mediated $\tilde{u}_2$ production dominates over direct $\tilde{u}_2$ production $pp \to \tilde{u}_2 \bar{\tilde{u}}_2 X$; the $\tilde{g}$ production rate is large, $\sigma(pp \to \tilde{g}\tilde{g}X) \approx 150$ fb, at LHC (14 TeV) and $B(\tilde{g} \to \tilde{u}_2 c/t)$ are large ($\sim 25\%$), which leads to a sizable $\tilde{u}_2$ production rate and hence sizable rates of QFV bosonic $\tilde{u}_2$ decay signals at LHC (14 TeV):

In scenario A, using the branching ratios in Table 1, we get the QFV signal rates
$\sigma(pp \to \tilde{g}\tilde{g}X \to t c \bar{c} \bar{c} h^0 E_T^{mis} X) = 8$ fb, $\sigma(pp \to \tilde{g}\tilde{g}X \to t t \bar{c} \bar{c} h^0 E_T^{mis} X) = 4$ fb,
where charge conjugated final states are included, $E_T^{mis}$ is missing transverse energy due to the two LSP neutralinos $\tilde{\chi}_1^0$, and X contains only the beam jets. For an integrated luminosity of 500 fb$^{-1}$ (ATLAS + CMS), we expect 4000 and 2000 events for these QFV signals, respectively.

In scenario C, using the branching ratios in Table 2, we get the QFV signal rates
$\sigma(pp \to \tilde{g}\tilde{g}X \to t c \bar{c} \bar{c} h^0 E_T^{mis} X) = 8.5$ fb, $\sigma(pp \to \tilde{g}\tilde{g}X \to t c \bar{c} \bar{c} Z^0 E_T^{mis} X) = 6.8$ fb,
$\sigma(pp \to \tilde{g}\tilde{g}X \to t t \bar{c} \bar{c} h^0 E_T^{mis} X) = 1.4$ fb, $\sigma(pp \to \tilde{g}\tilde{g}X \to t t \bar{c} \bar{c} Z^0 E_T^{mis} X) = 1.1$ fb,
where charge conjugated final states are included.

## 5. Conclusion

In case the Nature takes our scenarios or similar ones, then the "standard" MC-analysis for squark/gluino searches would lead to totally wrong results, such as wrong sparticle mass limits, and could even miss the true SUSY signals! Therefore, we encourage the ATLAS/CMS collaborations to perform the squark/gluino search by analysing these search modes steming from QFV bosonic squark decays in addition to the "standard" search modes steming from QFC fermionic squark decays at LHC (14 TeV) (not at LHC (7-8 TeV) where the QFV signal rates are too small in our scenarios). Moreover, the possible very large branching ratios of the QFV bosonic squark decays could eventually affect the MSSM parameter determination.